\documentclass[a4paper]{jpconf}
\usepackage{graphicx}

\begin{document}
\title{Revisiting the theoretical DBV (V777 Her) instability strip: 
the MLT theory of convection}

\author{A H C\'orsico$^1$, 
        L G Althaus$^1$, 
        M M Miller Bertolami$^1$ and \\
        E Garc\'\i a--Berro$^{2,3}$}  
\address{$^1$Facultad de Ciencias Astron\'omicas y Geof\'{\i}sicas,   
         Universidad Nacional de La Plata, 
         Paseo del Bosque S/N, (1900)   
         La Plata, 
         Argentina}  
\address{$^2$Departament de F\'\i sica Aplicada, 
         Escola Polit\`ecnica Superior de Castelldefels, 
         Universitat Polit\`ecnica de Catalunya,    
         Av. del Canal Ol\'\i mpic, s/n, 
         08860 Castelldefels, 
         Spain}  
\address{$^3$Institute for Space Studies of Catalonia,
         c/Gran Capit\`a 2--4, Edif. Nexus 104,   
         08034  Barcelona,  Spain}
\ead{acorsico@fcaglp.unlp.edu.ar} 

\begin{abstract}
We reexamine the theoretical  instability domain of pulsating DB white
dwarfs (DBV or V777 Her variables). We performed an extensive $g$-mode
nonadiabatic pulsation analysis  of DB evolutionary models considering
a wide  range of stellar  masses, for which the  complete evolutionary
stages  of their  progenitors  from the  ZAMS,  through the  thermally
pulsing AGB and born-again phases, the domain of the PG1159 stars, the
hot phase of  DO white dwarfs, and then the DB  white dwarf stage have
been  considered.  We  explicitly  account for  the  evolution of  the
chemical  abundance   distribution  due  to   time-dependent  chemical
diffusion  processes.    We  examine  the  impact   of  the  different
prescriptions of the MLT theory of convection and the effects of small
amounts of  H in  the almost  He-pure atmospheres of  DB stars  on the
precise location of  the theoretical blue edge of  the DBV instability
strip.
\end{abstract}

\section{Introduction}

Variable  DB white  dwarfs (also  called DBV  or V777  Her  stars) are
$g$-mode nonradial pulsators with periods ranging from $200$ to $1000$
s.    They  are   characterized  by   He-rich   atmospheres,  possibly
contaminated with small impurities of H.  Their pulsations are thought
to be triggered by the $\kappa-\gamma$ mechanism acting on the partial
ionization  of He  at  the base  of  the outer  convection zone.   The
observed instability strip of the DBV stars is located between $T_{\rm
eff} \approx 28\,400$ K and $T_{\rm eff} \approx 22\,500$ K (Winget \&
Kepler 2008). All of the published stability analysis of DB models ---
see,  for  instance,  Bradley   \&  Winget  (1994)  and  Beauchamp  et
al. (1999) --- clearly indicate a strong dependence of location of the
{\sl theoretical}  blue (hot) edge  of the DBV instability  strip with
the  convective efficiency  adopted  in the  envelope  of the  stellar
models.  At  present, there is no general  consensus between different
authors about what is the  right convective efficiency of DB models in
order to fit  the observed blue edge. On the  other hand, there exists
an additional  uncertainty in the definition of  the temperature scale
of  the  DB  stars, and  consequently  in  the  location of  the  {\sl
observed} blue edge, due to  the strong dependence with the convective
efficiency  adopted in  the model  atmosphere fits.   Finally, another
difficulty is  the uncertainty in the surface  parameters that results
from the presence  of undetectable amounts of H  in the atmospheres of
DB stars (Beauchamp et al. 1999).  In this work, we explore the impact
of different convective efficiencies and  of small amounts of H in the
atmospheres of  DB stars  on the precise  location of  the theoretical
blue edge of  the DBV instability strip.  We  employ the Mixing Length
Theory (MLT) of  convection to model the outer  convection zone of our
models.  We also  examine the  dependence of  the blue  edge  with the
stellar  mass. To this  end, we  perform fully  nonadiabatic pulsation
calculations of  up-to-date DB white dwarf evolutionary  models with a
wide range of stellar masses.

\begin{figure}[t]
\vskip 8cm
\includegraphics{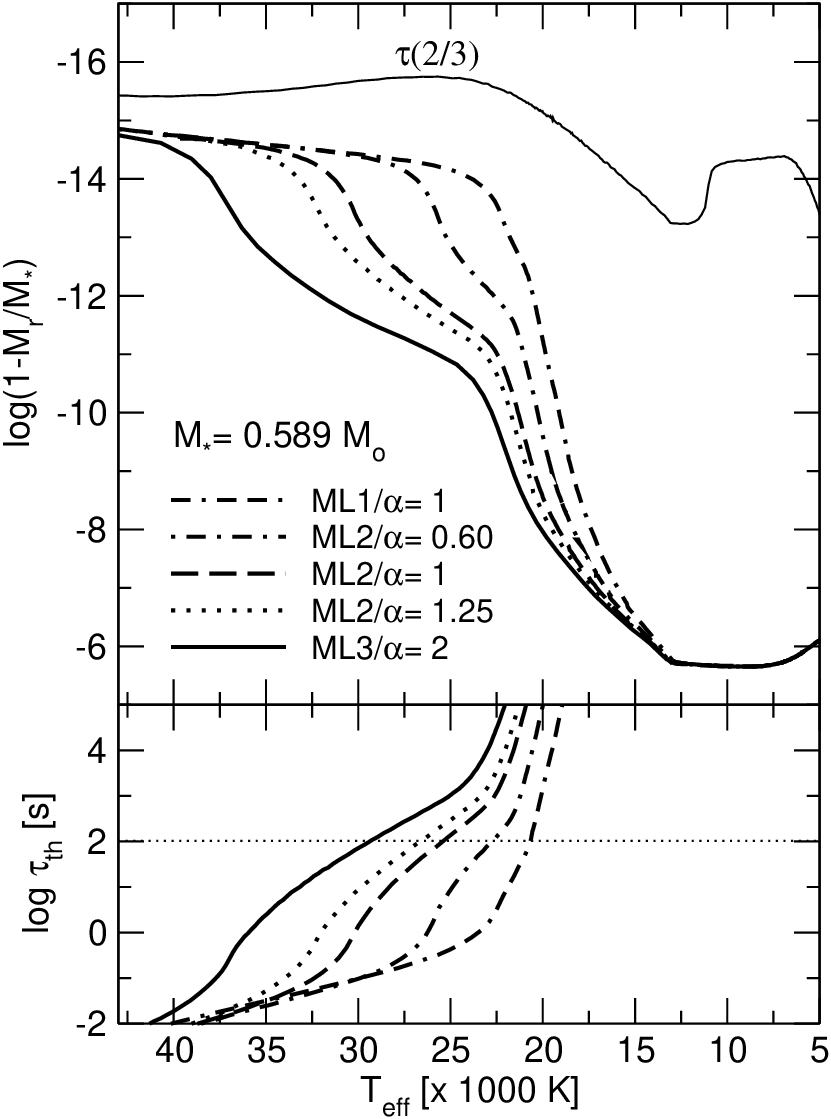}
\caption{\label{figure1} The  extent in  mass of the  outer convection
         zone in terms of $T_{\rm eff}$ for different prescriptions of
         the MLT  theory of convection  (upper panel), and the  run of
         the thermal  timescale at the  basis of the  outer convective
         zone (lower panel).}
\end{figure}

\section{Stellar structure, evolution and pulsation modeling}

We employ a  new set of fully evolutionary DB  white dwarf models that
descend from  the post-born again  PG1159 models (Miller  Bertolami \&
Althaus 2006).  Specifically, we  employ the {\tt LPCODE} evolutionary
code (Althaus et  al. 2005) to compute the  evolution of PG1159 models
(H-deficient,  He-, C-,  and O-dominated  atmospheres) towards  the DB
regime   (almost  He-pure   atmospheres)  taking   into   account  the
time-dependent  microscopic diffusion  of $^4$He,  $^{12}$C, $^{13}$C,
$^{14}$N, $^{16}$O  and $^{22}$Ne.  The  metallicity is assumed  to be
$Z=  0$  in  the  metal-free  He-rich  envelopes  and  $Z  \leq  0.02$
otherwise.  The He-rich envelopes of our models are characterized by a
double-layered chemical structure built  up by chemical diffusion.  We
perform nonadiabatic pulsation computations of nonradial $g$-modes for
$\ell= 1, 2$  with the code described in C\'orsico  et al. (2006).  We
analyze  the  pulsation  stability  of  DB  white  dwarf  models  with
effective temperatures between $35\,000$ K and $17\,000$ K and stellar
masses of $0.530, 0.542, 0.556, 0.565, 0.589, 0.609, 0.664, 0.741$ and
$0.870 M_{\odot}$.  We model convection  in the envelope of our models
in  the  framework  of  the   MLT  theory.  We  employ  the  following
prescriptions  of   the  MLT,  in  order   of  increasing  efficiency:
ML1$/\alpha=  1$, ML2$/\alpha=  0.60$,  ML2$/\alpha= 1$,  ML2$/\alpha=
1.25$, and  ML3$/\alpha= 2$, $\alpha$ being  the characteristic mixing
length (Tassoul et al. 1990).

\begin{figure}[t]
\vskip 5cm
\includegraphics{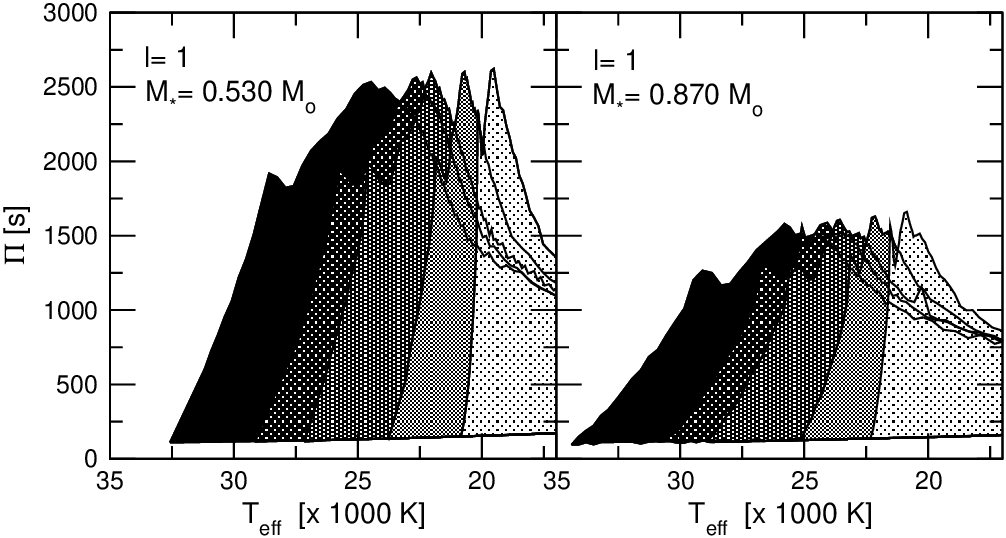}
\caption{\label{figure2} The  instability domains of  dipole modes for
         different MLT prescriptions (from left to right: ML3$/\alpha=
         2$, ML2$/\alpha= 1.25$,  ML2$/\alpha= 1$, ML2$/\alpha= 0.60$,
         and ML1$/\alpha= 1$).}
\end{figure}

\section{Results}

The evolution and extent of the  outer convective zone of our DB white
dwarf sequences during the DBV instability strip is shown in the upper
panel  of  Fig. \ref{figure1}  for  the  different  treatments of  the
MLT. Note that  the base of the outer  convective zone notably deepens
with cooling,  being this effect more  pronounced in the  case of more
efficient MLT prescriptions in the proximity of the blue edge ($T_{\rm
eff} \approx  29\,000$ K).  In the  lower panel we display  the run of
the thermal timescale  at the basis of the  outer convective zone.  DB
models  are expected  to became  $g$-mode pulsationally  unstable when
$\tau_{\rm  th} \approx  100$  s.   This would  happen  first for  the
highest  convective  efficiency  (ML3$/\alpha=  2$)  models.  This  is
confirmed through full nonadiabatic computations.

Fig.  \ref{figure2}  displays the  instability domains in  the $T_{\rm
eff}-\Pi$ diagram, corresponding to  the sequence of $0.530 M_{\odot}$
(left)  and  the  sequence  of  $0.870  M_{\odot}$  (right),  for  the
different  MLT prescriptions.  On the  basis  of this  figure, we  can
outline  several trends:  (1)  there  is a  strong  dependence of  the
longest excited periods  with the stellar mass, being  larger for less
massive models; (2)  the shorter excited periods, on  the contrary, do
not exhibit any dependence with $M_*$; (3) regarding convection, there
is no dependence of the longest (nor the shorter) excited periods with
the particular MLT prescription adopted; (4) the blue edge is strongly
sensible to the convective efficiency, being much hotter for efficient
convection (that is, for ML3 and ML2$/\alpha= 1.25$); and (5) the blue
edge is hotter for more massive DB models.

\begin{figure}[t]
\vskip 6.9cm
\includegraphics{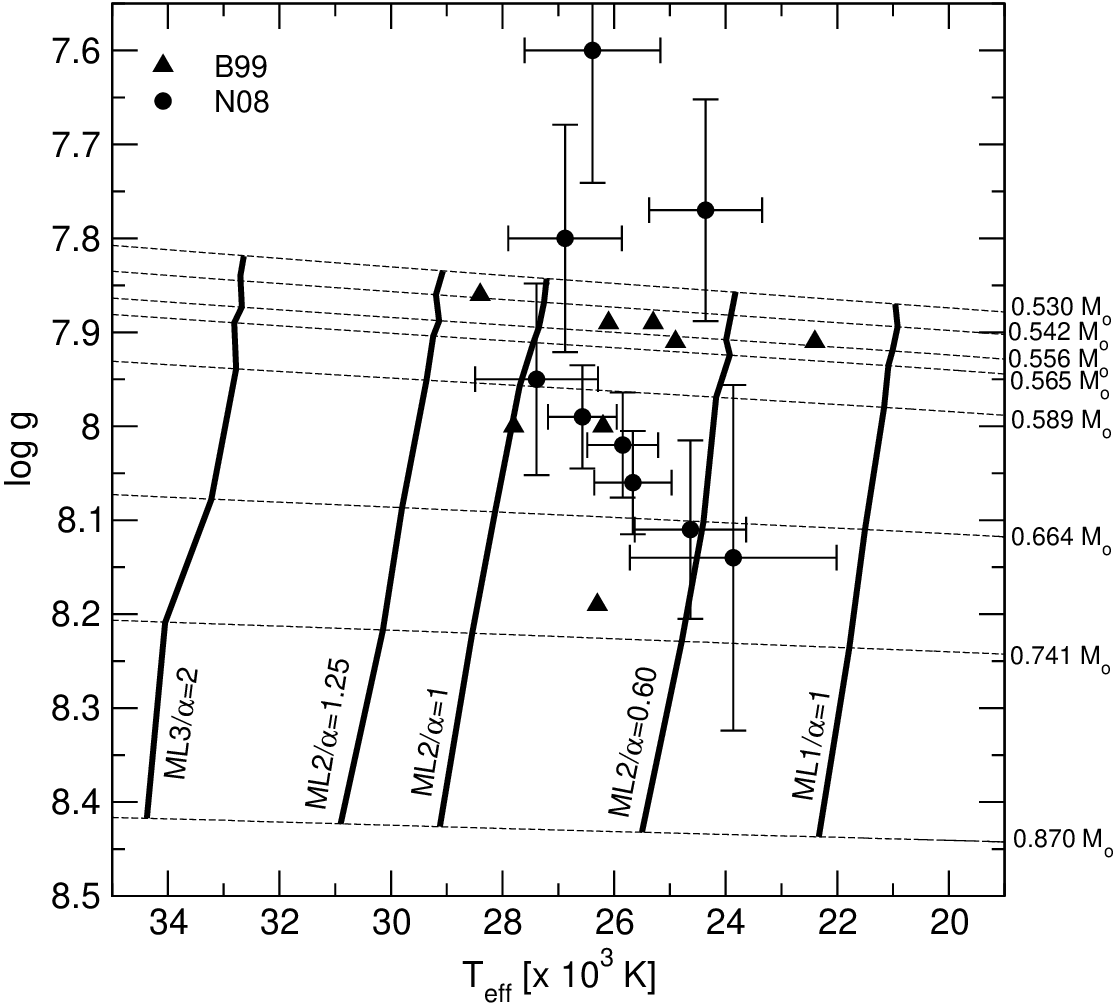}
\caption{\label{figure3} The  blue edge  of the DBV  instability strip
         for different recipes of the MLT.}
\end{figure}

Now,  we examine  how our  theoretical blue  edges of  the instability
strip compare  with the  location of the  observed pulsating  DB white
dwarfs.  We  consider the eight  DBV stars considered in  Beauchamp et
al. (1999) , plus the nine DBVs recently discovered in the SDSS (Nitta
et al. 2009). We consider only  the case of atmospheres devoided of H.
Fig.   \ref{figure3}  shows  our   complete  set  of  DB  evolutionary
sequences (dashed  lines) on the  $T_{\rm eff}-\log g$  diagram, along
with the  blue edge  for the different  MLT prescriptions.   Note that
there is a clear dependence of the blue edge with the stellar mass. It
is apparent that  only the ML2$/\alpha= 1.25$ prescription  is able to
account for the location of all known DBVs.

\section{The effects of H on the location of the blue edge}

At present, it is a well known observational fact that the atmospheres
of DB white  dwarf stars exhibit small traces  of H, attributed mainly
to  accretion  from  the  ISM.   A  modest  accretion  rate  of  about
$10^{-19}-10^{-21}  M_{\odot}/$yr  would  be  enough  to  explain  the
presence of  H in the  DBs. The effective  temperature of DB  stars as
derived from  model atmospheres that contain H  are considerably lower
than in  the cases  in which  H is neglected  (Beauchamp et  al. 1999;
Castanheira et al. 2006; Voss et al. 2007). It is expected also that a
non-negligible shift  of the theoretical blue edge  should result when
the He-rich envelopes of  the equilibrium models are contaminated with
small impurities of  H.  To test this possibility,  we have restricted
ourselves to the  sequences of $M_*= 0.530 M_{\odot}$  and $M_*= 0.741
M_{\odot}$.   For  these two  values  of  the  stellar mass,  we  have
explored the  cases in  which $X_{\rm H}=  0.0001, 0.001$  and $0.01$,
corresponding to $\log(n_{\rm H}/n_{\rm He})= \log (4 X_{\rm H}/X_{\rm
He})= -3.4, -2.4$ and $-1.44$, respectively.  We found that if $X_{\rm
H}= 0.0001$,  there is no appreciable  effects on the  location of the
blue edge.  However, if $X_{\rm H}= 0.001$ the blue edge is shifted to
lower effective temperatures by $\sim 800$ K (for ML3$/\alpha= 2$) and
$\sim 300$ K (for ML1$/\alpha=  1$).  Finally, if $X_{\rm H}= 0.01$ we
obtain a shift of $\sim 3000$  K (for ML3$/\alpha= 2$) and $\sim 1200$
K (for ML1$/\alpha= 1$).

\section{Conclusions}
\label{conclusions}

The  results of  the present  work  indicate that  the best  agreement
between the location of the  DBV stars previously studied in Beauchamp
et al.  (1999) with  surface parameters from H-free model atmospheres,
and  the predictions  of  the nonadiabatic  pulsation calculations  is
provided  by the  case of  ML2$/\alpha= 1.25$.   Notably,  the $T_{\rm
eff}$ and $\log  g$ previously derived in Beauchamp  et al. (1999) are
obtained with  model atmospheres that assume  ML2$/\alpha= 1.25$.  Our
ML2$/\alpha=  1.25$ blue  edge is  consistent also  with the  SDSS DBV
stars  reported  in  Nitta  et  al.  (2009),  although  their  surface
parameters are derived from model atmospheres with ML2$/\alpha= 0.60$.
The present work  constitutes the first phase of  a thorough pulsation
study (adiabatic and nonadiabatic) of DBV  stars on the basis of a new
generation of fully evolutionary DB models.

\ack

We thank S. O. Kepler for providing us the $T_{\rm  eff}$ and $\log g$
values of the  recently discovered SDSS DBV stars.   Part of this work
was  supported  by   PIP  6521  grant  from  CONICET,   by  MEC  grant
AYA05--08013--C03--01, by  the European Union  FEDER funds and  by the
AGAUR.

\section*{References}

\begin{thereferences}

\item Althaus L G Serenelli A M Panei et al. 2005 {\sl A\&A} {\bf 435}
      631
\item Beauchamp  A Wesemael F  Bergeron P et  al. 1999 {\sl  ApJ} {\bf
      516} 887
\item Bradley P A \& Winget D E 1994 {\sl ApJ} {\bf 421} 236
\item Castanheira B G Kepler S O Handler G \& Koester D 2006 {\sl A\&A} 
      {\bf 450} 331
\item C\'orsico  A H  Althaus L G  \& Miller  Bertolami M M  2006 {\sl
      A\&A} {\bf 458} 259
\item Miller  Bertolami M M \& Althaus  L G 2006 {\sl  A\&A} {\bf 454}
      845
\item Nitta A Kleinman S J Krzesinski J et al.  2009 {\sl J. of Phys.:
      Conf. Ser.}, in press
\item Tassoul M Fontaine G \& Winget D E 1990 {\sl ApJS} {\bf 72} 335
\item Voss  B Koester D  Napiwotzki R Christlieb  N \& Reimers  D 2007
      {\sl A\&A} {\bf 470} 1079
\item Winget D E \& Kepler S O 2008 {\sl ARA\&A} {\bf 46} 157

\end{thereferences}

\end{document}